\documentclass{aa}
\usepackage{epsf}

\def\R{{\cal R}}
\def\bx{\mathbf{x}}
\def\bt{\mathbf{\theta}}

\def\beq#1{\begin{equation}\label{#1}}
\def\eeq{\end{equation}}
\def\beqa#1{\begin{eqnarray}\label{#1}}
\def\eeqa{\end{eqnarray}}

\def\myfrac#1#2{\left(\frac{#1}{#2}\right)}
\def\comment#1{\relax}

\def\spose#1{\hbox to 0pt{#1\hss}}
\def\simlt{\mathrel{\spose{\lower 3pt\hbox{$\mathchar"218$}}
     \raise 2.0pt\hbox{$\mathchar"13C$}}}
\def\simgt{\mathrel{\spose{\lower 3pt\hbox{$\mathchar"218$}}
     \raise 2.0pt\hbox{$\mathchar"13E$}}}
\def\simpropto{\mathrel{\spose{\lower 3pt\hbox{$\mathchar"218$}}
     \raise 2.0pt\hbox{$\propto$}}}

\makeatletter
\def\eqalign#1{\null\,\vcenter{\openup\jot\m@th
  \ialign{\strut\hfil$\displaystyle{##}$&$\displaystyle{{}##}$\hfil
      \crcr#1\crcr}}\,}
\def\eqalignleft#1{\null\,\vcenter{\openup\jot\m@th
  \ialign{\strut$\displaystyle{##}$\hfil&$\displaystyle{{}##}$\hfil
      \crcr#1\crcr}}\,}
\makeatother

\def\R{{\cal R}}
\def\L{{\cal L}}

\begin{document}
\setcounter{footnote}{1}

\thesaurus{12       
     (02.07.2; 
      08.02.1)} 
     
\title{Fluctuations of Gravitational Wave Noise from Unresolved 
Extragalactic Sources}
\author{
D.I.Kosenko
K.A.~Postnov}

\institute{
Faculty of Physics and 
Sternberg Astronomical Institute, Moscow University,
                119899 Moscow, Russia}
\date{Received ... 1999, accepted ..., 1999}
\maketitle
\markboth{D.Kosenko \& K.Postnov. Fluctuations of Extragalactic GW Noise }{ ...}

\begin{abstract} Angular fluctuations
of stochastic 
gravitational wave backgrounds (GWB) produced by extragalactic 
astrophysical sources are calculated. 
The angular properties of such backgrounds 
are determined by the large scale structure of Universe (galaxy 
clustering). The evolution of star formation rate with redshift 
is taken into account. Fluctuations of the 
metric strain amplitude associated with such noises 
at angular scales of about one degree are found to be of order 
5-20\% slowly growing toward smaller angular scales.
This feature can be potentially used to separate astrophysical GWB from 
cosmological ones in future experiments.


\end{abstract}

\section{Introduction}

With the advent of detectors of gravitational waves (GW),  which are 
currently under construction, a totally new possibility to study
different astronomical objects opens up (Thorne 1988). In addition to 
"classic" sources of GW like merging compact binaries or 
rapidly rotating hot neutron stars etc., which will be studied  
at frequencies $\sim 10-1000$ Hz, there should exist a specific 
cosmological background (noise) covering a very wide frequency band
from $\sim 10^{-18}$ to $10^5$ Hz. The cosmological background 
should bear imprints of the physical
processes in the very early Universe (Grishchuk 1988, 1997 and references 
therein). The primordial GW
background, which originates from vacuum fluctuations 
parametrically amplified by the very expansion of the Universe, has 
a power-law spectrum spanning a very wide frequency range  
(Starobinsky 1979, Rubakov et al. 1982, 
Grishchuk 1988, 1997). The prospects for its detection appear to be the most 
favorable at the low frequency band ($10^{-3}-10^{-1}$ Hz) which
will be covered by LISA space interferometer (Larson et al. 1999).   

At some frequencies, however, unresolved 
binary stars within our own Galaxy or beyond provide an
important contribution in the LISA frequency band (Bender and Hils 1997, 
Postnov and Prokhorov
1998, Kosenko and Postnov 1998). As shown in these papers, the 
stochastic signal from unresolved 
merging white dwarf binaries 
dominates the LISA sensitivity curve up to $\sim 3\times 10^{-3}$
Hz. At higher frequencies 
extragalactic merging white dwarf binaries contribute at a level
roughly 10 times smaller, which is below the planned LISA sensitivity
at these frequencies. This fact allows search for 
detection of the primordial GW backgrounds by LISA (Grishchuk 1997). 

However, some inherent uncertainties (e.g. in the galactic
rate of binary white dwarf mergings) are present in the calculations 
so the astrophysical  backgrounds can turn out to be higher than expected. 
Since astrophysical GW backgrounds are considered
as an additional noise contributing to the intrinsic noise 
of the detector, as much as possible of their properties 
at all frequencies should be known in advance.  

What are specific features of astrophysical GW backgrounds? Clearly, 
those related to the galactic sources should follow the distribution of
stars inside the Milky Way (Hils et al. 1989, 
Lipunov et al. 1995). As all the detectors (ground-based or space-born)
should rotate with respect to the galactic plane, 
the signal modulation has been used as an 
advantage to detect them (Giazotto et al. 1997, Giampieri and Polnarev 
1997). As for the backgrounds of extragalactic origin, only 
amplitudes at different frequencies from various sources have
been computed so far (Kosenko and Postnov 1998, Ferrari et al. 1999a,b).

The purpose of the present paper is to study angular properties of
the GW noise produced by extragalactic astrophysical sources at the 
degree scales. 
As most of these source must reside in galaxies (only a tiny fraction of 
binaries or GW-emitting neutron stars is expected to be in the
intergalactic space), the GW background should have distinctive 
angular correlation
properties reflecting the large scale structure (LSS) of the Universe.
This is exactly what we observe in electromagnetic radiation as 
fluctuations of, for example, IR background observed by COBE 
(Kashlinsky et al. 1999).

\section{Imprint of the LSS in the astrophysical GWB}

The angular fluctuations of intensity in any astrophysical GWB
should correlate with those of the projected number of galaxies, so 
first we wish to remind the reader the
well-developed technique which is used 
for studies of angular properties of galactic 
counts on the sky. There is a lot of specially dedicated literature 
on this subject (see e.g. Peebles 1980, 1993). 
Before to proceed further, let us make a simple estimate. Suppose 
we have sources (galaxies) randomly distributed in space. What 
is the rms fluctuation of the number of galaxies within a small solid angle
$\delta\Omega=\pi\theta^2\ll 4\pi$? With the average space density of
galaxies $n_G\simeq 0.01$ Mpc$^{-3}$ and neglecting cosmological effects
for a while we would get 
\beqa{}
\frac{\delta N(\theta)}{N}=\frac{1}{\sqrt{N(\theta)}}
&\sim& 
\frac{1}{\sqrt{n_G}}\frac{1}{\theta}\,\left(\frac{c}{H_0}\right)^{-3/2} 
\\&\sim& \nonumber
3\times 10^{-2}\left(\frac{\theta}{0.01}\right)^{-1}\,,
\eeqa       
(here $H_0$ is the Hubble constant)
while galactic counts demonstrate a much higher value $\delta N/N\sim 0.5$ 
for linear scales $l=30$ Mpc (i.e. $\theta\sim 0.01$ for the most distant
galaxies) (Peebles 1993). So clearly we should take into account 
the correlation properties of galaxies and galactic clusters. 

To account for the non-Poissonian properties of galactic distribution in
space, the correlation functions are used. For Gaussian fluctuations 
only two-point correlation function $\xi(\vec r)$ or its Fourier
transform (power spectrum) $P_3(\vec k)$ would be sufficient. 
Taking $P_3(k)$ as derived from LSS studies, we then can calculate
the 2-dimensional correlation function $C(\theta)$ of the projected 
distribution of galaxies on the sky or, 
equivalently, the 2-dimensional power spectrum $P_2(q)$. This is 
the last quantity that we actually need since the rms fluctuation of 
the energy flux per unit logarithmic frequency interval in a 
given direction is directly related to $P_2(q)$ 
(Kashlinsky et al. 1999 and below).

\subsection{Correlation functions and power spectrum}

The brightness of a GW background
can be characterized by the energy flux coming from a given direction 
within a solid angle
$\Delta\Omega$ (Thorne 1988) 
\beq{}
f\frac{dE(\mathbf{x})}
{dtdSdfd\Omega}\Delta\Omega=\int\limits_{\Delta\Omega} fI_fd\Omega
\eeq
where $\mathbf{x}$ is the two dimensional coordinate across the sky. 
The integration over all sky yields the familiar value 
$\Omega_{GW}\rho_{cr}c^2$, the energy density per unitary 
logarithmic frequency interval in units of the critical energy density
to close the Universe, which characterizes the isotropic stochastic
GWB. To study angular properties of the noise we
shall consider the flux from given direction  
$F(\mathbf{x})\equiv fI_f$ itself.
We shall assume some ideal GW detector with a beam-like sensitivity 
diagram, which is of course far from realistic
ground-based LIGO-like or
spaceborn LISA-like interferometers. However, in this paper 
we will not 
discuss the observability of GW backgrounds.   

The fluctuation in the GW flux arrived at the detector from 
a given direction on the sky is $\delta
F(\mathbf{x})=F(\mathbf{x})-\langle F\rangle$, where $\langle\ldots\rangle$
denotes ensemble averaging. The Fourier transform of the fluctuation
is $\delta F(\bt)=1/(2\pi)^2\int \delta F_q \exp(-i\mathbf{q}\bt)
d^2\mathbf{q}$.

The projected 2-dimensional correlation function represents the first
non-trivial moment of the probability distribution function of $\delta
F(\bx)$:  $C(\bt)=\langle \delta F(\bx+\bt)F(\bx)\rangle$. The two
dimensional power spectrum is by definition 
$P_2(q)=\langle |\delta F_q|^2\rangle$. We shall assume the phases
to be random and the distribution of the flux field to be Gaussian so
that the power spectrum is just the Fourier transform of the correlation 
function:

\beq{}
C(\theta)=\frac{1}{2\pi}\int_0^\infty P_2(q) J_0(q\theta)qdq\,,
\eeq
\beq{P2}
P_2(q)=\int_0^\infty C(\theta) J_0(q\theta)\theta d\theta \,.
\eeq
Here $J_0$ is the zero-order cylindrical Bessel function.
    
The mean square fluctuation of the flux on the detector 
within a finite solid angle $\Delta\Omega$
subtended by the angle $\vartheta$ across the sky 
is zero-lag correlation signal

\beq{}
\langle(\delta F)^2\rangle_\vartheta =\frac{1}{2\pi}\int_0^\infty
P_2(q)W(q\vartheta)qdq
\eeq  
where $W$ is the window function of the detector. For example,
for a top-hat beam $\langle(\delta F)^2\rangle_\vartheta
\sim (1/2\pi)q^2P_2(q)|_{q\sim \pi/2\vartheta}$ and the values 
of $1/q$ correspond to fluctuations of angular size $\sim \pi/q$. 

The GWB flux and its angular properties measured in projection on the
celestial sphere should reflect 3-dimensional structure of the Universe and
the change of GW emission rate with redshift $z$. The LSS can be taken into 
account by the 3-dimensional correlation function $\xi(r)$  or its
3-dimensional power spectrum $P_3(k)$, which for isotropic case 
relates to $\xi(r)$ through the equation

\beq{}
\xi(r)=\frac{1}{2\pi^2}\int_0^\infty P_3(k)j_0(kr)k^2dk
\eeq 
(here $j_0$ is the zero-order spherical Bessel function).

The projected correlation function of cosmic GWB  $C(\bt)$ is expressed 
through two-point correlation function of the galaxy distribution
$\xi(r_{12})$ and
the rate of GW emission $dF/dz$ via the Limber equation (Limber, 1953):

\beq{Limber}
C(\bt)=\int\myfrac{dF}{dz_1}\myfrac{dF}{dz_2}\xi(r_{12},z)dz_1dz_2
\eeq
Substituting this equation in the limit of small angles $\theta\ll 1$
for Friedman-Robertson-Walker metrics 
into Eq. (\ref{P2}) one gets (Kashlinsky et al. 1999)
 
\beq{P2q}
P_2(q)=\int_0^\infty\myfrac{dF}{dz}^2\frac{P_3(\frac{q}{d_A(z)},z)}
{d_A^2(z)c\frac{dt}{dz}}dz
\eeq 
where $d_A=d_m/(1+z)$ is the angular distance, $d_m$ is the metric 
distance. For degree angular scales 
of interest here the linear approximation of galactic 
clustering can be used $P_3(k,z)\simeq P_3(k,0)\times \Psi(z)$, where
$\Psi(z)$ describes the evolution of the clustering. On linear scales
$\Psi(z)=(1+z)^{-1}$ if $\Omega=1$ (Peebles, 1980).

In our analysis we use the 3D spectrum $P_3(k)$ as derived by 
Einasto et al. (1999) from a thorough analysis of different 
LSS studies. The mean spectrum of galaxies shows a power-law behavior
at small and large $k$ with a maximum $\sim 10^4[h^{-3}\hbox{Mpc}^3]$
at $k\approx 5\times 10^{-2}\,h\,\hbox{Mpc}^{-1}$.   

The analysis of this formula (Kashlinsky et al. (1999) and references
therein) shows that the relative fluctuations of the flux are 
$\delta F_{rms}\sim \sqrt{H_0/c k^2P_3(k)}\sim 5\%-10\%$ and weakly 
dependent on the cosmological model at angular scales of order one degree. 
However, strong dependence on redshift of the flux rate 
requires more accurate calculations.

\subsection{GWB flux}

The GW energy  
emitted by the population of 
some sources in a galaxy at frequency $f$
per 
unit logarithmic frequency interval $d\ln f$  
in the rest-frame of the galaxy can be calculated 
through the rate of GW-producing events $\R$ (e.g. the rate of
binary WD mergings or supernova explosions) in the galaxy. Under the 
stationary conditions we have 
(Kosenko and Postnov 1998)

\beq{}
\frac{dE}{dtd\ln f}=\R \frac{E(f)}{dt}\left(\frac{d\ln
f}{dt}\right)^{-1}_{GW}
\eeq 
where $E(f)$ is the energy which is being carried away by gravitational
waves from the typical source (e.g. for two point masses orbiting each other 
it is just the orbital binding energy). If $E(f)\propto f^\alpha$
and GW emission is the only dissipative mechanism 

\beq{dE}
\frac{dE}{dtd\ln f}=\alpha\R E(f)
\eeq

The comoving luminosity at proper frequency $f'$ 
per unit logarithmic frequency interval 
produced by sources in galaxies at redshift
$z$ from the redshift interval $dz$ from unit solid angle 

\beq{dL}
\frac{d\L(f',z)}{dz}=n(z)\frac{\delta V(z)}{dz}\frac{dE(f')}{dtd\ln f'}
\eeq
where $n(z)$ is the space density of galaxies, $\delta V(z)$ is the proper
volume element. 
We assume no new galaxies to create since their formation 
so that $n(z)=n_G(1+z)^3$ and
$n_G=0.013(\Omega_b/0.005)h_{100}^2$~Mpc$^{-3}$
is the present-day density of galaxies normalized to the amount of baryons
comprised in stars (in terms of the critical energy density $\rho_{cr}c^2$;
$h_{100}\equiv H_0/100$(km/s/Mpc) is the Hubble constant). The strong star
formation rate evolution is taken into account through the evolution of 
the event rate with redshift $\R(z)$ (Kosenko and Postnov 1998). We use
the parametrization of Rowan-Robinson (1999) for the star formation rate 
history SFR$(z)$ as derived from optical, UV, and IR-observations, and 
normalized to unity at $z=0$. 
The rate of particular events is thus
$\R(z)=\R_G\int\limits_{z_*}^0\hbox{SFR}(z')G(z-z')dz'$ where 
$\R_G$ is the present-day event rate per galaxy,
$G(z-z')$ is the redshift 
dependence of the rate after a $\delta$-function-like star formation burst.
This is important especially for binary white dwarf coalescences
because these are delayed typically by $\sim 10^9$ years since their 
formation.
For events which relate to massive star evolution,  
like supernova explosions,  we have $G(z-z')\approx \delta(z-z')$ and
the change in the rate of events with redshift simply 
follows the star formation rate evolution 
$\R(z)=\R_G\times \hbox{SFR}(z)$.         

The proper volume element is (Peebles 1993)
\beq{dV}
\frac{\delta V(z)}{dz}= c\frac{dt}{dz}\frac{d_m(z)^2}{(1+z)^2}
\eeq 
with the metric distance $d_m$ and $dt/dz$ being the functions of 
the cosmological model (Carroll et al. 1992). In our calculations we use 
the standard flat universe without cosmological constant 
and a $\Lambda$-term dominated cosmological model with 
$\Omega_\Lambda=0.7$, $\Omega_m=0.3$.

Combining equations (\ref{dE}), (\ref{dL}), and (\ref{dV}) together we
arrive at 
\beq{}
\frac{d\L(f',z)}{dz}=c\frac{dt}{dz}d_m^2(z)(1+z)n_G\R(z)E(f')\alpha  
\eeq
so the contribution to the total GW flux from the redshift interval $dz$ from
unit solid angle which is observed today by a detector with band-width $df$
centered at frequency $f=f'/(1+z)$ is

\beqa{}\nonumber
\frac{dF(f)}{dz}
&\equiv&
\frac{dE(f)}{dt\;dS\;d\ln f\,dz}
\\[.5ex]&=&\label{dF}
\myfrac{d\L(f(1+z),z)}{dz}\frac{1}{4\pi d_L^2}
\\[.5ex]&=&\nonumber
\alpha\R_G(z)n_G E(f)\frac{(1+z)^{\alpha-1}}{4\pi}\frac{cdt}{dz}
\eeqa 
where we used the luminosity distance definition $d_L=d_m(1+z)$. 
Note that no 
additional factor $(1+z)$ accounting for the 
change of frequency interval appears in the numerator
since we are working with unitary logarithmic 
frequency interval. 

Now we are in the position to calculate relative fluctuations $\delta F/F$ 
of any GWB produced by astrophysical sources associated with 
galaxies  
within a given solid angle $\theta\ll 1$. From Eqs. (\ref{P2q}) and (\ref{dF})
we derive
\if!=
\beq{dFF}
\frac{\delta F(q)}{F}=\frac{\left(
\int\limits_0^{z_*} \myfrac{\R(z)}{\R_G}^2(1+z)^{2\alpha-
2}\Psi^2(z)\myfrac{cdt}
{dz}\myfrac{q}{d_A(z)}^2P_3\myfrac{q}{d_A(z)}
dz\right)^{1/2}}{\int\limits_0^{z_*} 
\myfrac{\R(z)}{\R_G}(1+z)^{\alpha-1}
\myfrac{cdt}{dz}dz}
\eeq
\fi
\beqa{}\nonumber
\frac{\delta F(q)}{F}&=&
\Biggl(\int\limits_0^{z_*} 
\myfrac{\R(z)}{\R_G}^2(1+z)^{2\alpha-2}\Psi^2(z)\myfrac{c\;dt}
{dz}
\times \\ 
&&\hphantom{\Biggl(\int\limits_0^{z_*}}
\times \label{dFF}
\myfrac{q}{d_A(z)}^2P_3\myfrac{q}{d_A(z)}
dz\Biggr)^{1/2}
 \Big/ \\ 
&\Big/& \nonumber
\int\limits_0^{z_*} \myfrac{\R(z)}{\R_G}(1+z)^{\alpha-1}
\myfrac{cdt}{dz}dz
\eeqa
Here the redshift $z_*$ corresponds to the beginning of 
star formation in the Universe. In our calculations we assumed $z_*=10$ 
(in fact, the exact value is of minor importance since sources at 
small and moderate redshifts mostly contribute to the flux).

For stochastic GWB the flux per unit logarithmic frequency interval  
is related to the dimensionless strain amplitude $h$ as

\beq{Sh}
h=\myfrac{2G}{\pi c^3f^2}^{1/2}F(f)^{1/2}
\eeq
so at a given frequency fluctuations in the strain amplitude
$\delta h/h=(1/2)(\delta F/F)$.

Note that the value of fluctuations depends on the specific source 
type only through the dependence of energy carried away by gravitational
waves on frequency (index $\alpha$ for the power-law dependence)
and the dimensionless change of event rate with redshift 
$\myfrac{\R(z)}{\R_G}$. The dependence on frequency can appear only
through the change in spectral index $\alpha$ 
with frequency. To ensure that the spectral shape has no effect 
on the relative fluctuations, 
observations should be performed at a frequency which 
is sufficiently (by an order of magnitude) lower than the high-frequency
cut-off of the comoving power-law spectrum of the background.

\subsection{Specific examples}

As specific examples we consider 
GWB produced by extragalacting merging
white dwarfs and rapidly rotating hot neutron stars. In the first case
the energy carried away by gravitational radiation is the orbital energy 
of binary white dwarfs whose evolution is driven by gravitation wave 
emission. In the LISA sensitivity frequency band $10^{-4}-10^{-1}$ Hz 
we have $E(f)\sim f^{2/3}$, $\alpha=2/3$. 
A white dwarf binary system reaches the orbital 
frequency of the pre-merging stage 
$10^{-2}$~Hz typically $\sim 10^9$ years after the formation,
so the dependence of the rate of events on redshift goes smoother 
than star formation rate history SFR$(z)$ in the Universe.   
The isotropic stochastic GWB 
produced by extragalactic binary white dwarfs with account of the 
star formation rate evolution in the Universe was calculated 
by Kosenko \& Postnov (1998). At the frequency $0.01$~Hz its
level is $h_{wd} \sim 5\times 10^{-21}$.

\begin{figure}
\epsfxsize=\hsize
\epsfbox{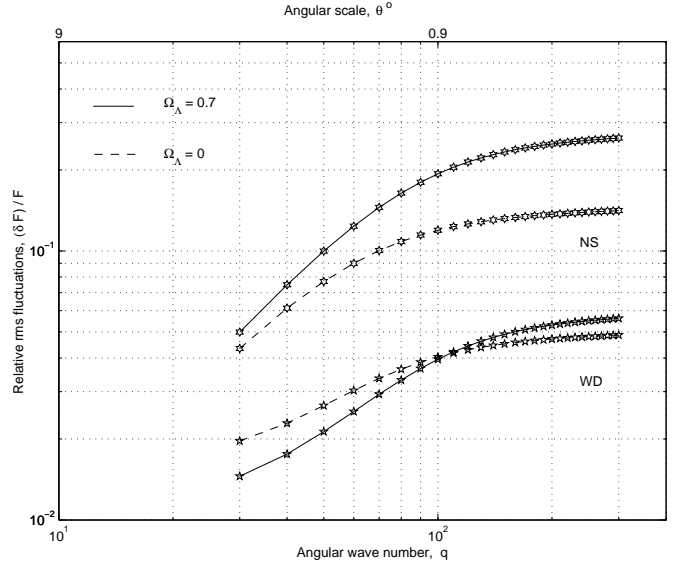}
\caption{Relative rms fluctuation of the gravitational wave 
flux from unresolved extragalactic 
astrophysical sources at different  angular scales $\theta=\pi/q$. 
Shown are fluctuations of the 
backgrounds produced by hot rapidly rotating neutron stars
at frequency 100 Hz 
(upper curves) and by coalescing binary white dwarfs
at frequency $10^{-2}$ Hz (lower curves).
Calculations for flat FRW universes without cosmological constant
($\Omega_L=0$) and with $\Omega_L=0.7$.}
\label{ref_fluc}
\end{figure}

For hot young neutron stars
the emission of gravitational waves can be driven by 
r-mode instability (Lindblom et al. 1998, Owen et al. 1999). 
The energy carried away by 
gravitational waves is the rotational energy of the neutron star $E_r$
and  $E_r(f)\sim f^2$ (i.e. $\alpha=2$) 
within the frequency band $\sim 120\, \hbox{Hz}<f<(2/3\pi)\Omega_K$, 
where $\Omega_K$ is the Kepler frequency at which mass shedding at the
stellar equator makes the star unstable. 
The isotropic stochastic GWB produced by hot NS was studied by
Ferrari et al. (1999b). Hot neutron stars are formed in the core collapse 
supernova explosion events in the end of evolution of massive stars, so no 
deviation from star formation dependence on redshift for their rate is
expected. The isotropic background level at 100 Hz was estimated to
be $h\sim 3\times 10^{-25}$.

The relative fluctuations of these backgrounds at angular scales
$\theta=\pi/q$ are shown in Fig. \ref{ref_fluc} as a function of
wave numbers $q$. It is seen that at $\theta\simlt 1$ degree
($q \simgt 100$), where the (quasi)linear regime of the galaxy clustering
evolution with redshift is expected so that we can use $\Psi(z)=(1+z)^{-1}$,
the fluctuations amount to 5\% for coalescing white dwarfs and 10-20\%
for hot neutron stars. At larger scales (smaller $q$) the relative
fluctuations slowly decrease. The calculations were performed for two
cosmological models: a flat FRW universe 
with zero cosmological constant and with
$\Omega_\Lambda=0.7$. In the case of hot neutron stars the level of
fluctuations increases in the $\Lambda$-dominated Universe at all scales
considered, while in the case of coalescing white dwarfs the curves for two
cosmological models intersect at $q\sim 110$ (angular scales $\theta\sim
1.5$ degrees).

Note that at smaller angular scales (larger $q$) 
the non-linear evolution of
LSS should be taken into account, so the present calculations
cannot be applied.

\section{Conclusion}

We have calculated the expected level of angular fluctuations of 
cosmic stochastic gravitational wave backgrounds produced by
populations of astrophysical sources associated with galaxies. 
The dependence of the source formation rate on redshift is 
taken into account using global star formation history 
in the Universe. The relative rms fluctuations of the 
GW flux does not depend on the specific source formation 
rate per galaxy, 
only on its evolution with redshift.
The level of relative flux 
fluctuations at angular scales $\theta \simlt 1$ degree
is found to be $\sim 5\%$ for coalescing white dwarfs with 
insignificant dependence on the cosmological model assumed, 
and $10-25\%$ for hot neutron stars, with stronger fluctuations
for flat cosmological constant dominated Universe.
Angular dependence of these fluctuations is 
a distinctive feature of such GW backgrounds and can be used
to discriminate between astrophysical and relic cosmological 
stochastic noises in future experiments.

\vskip\baselineskip

We thank Prof. A.V.Zasov for useful notes.  
The work was partially supported by Russian Fund for Basic Research
through Grant No 98-02-16801, 
the grant University of Russia 5559, 
and Royal Society Grant RCPX219.

\end{document}